# HIGH ENERGY COOLING

Valeri Lebedev[1], JINR, Russia 141980

Abstract: The paper considers methods of particle cooling mostly concentrating on cooling of high energy heavy particles in the high energy colliders. Presently, there are two major methods of the cooling the electron cooling and stochastic cooling. The latter can be additionally separated on the microwave stochastic cooling, the optical stochastic cooling (OSC) and the coherent electron cooling (CEC). OSC and CEC are essentially extensions of microwave stochastic cooling operating in 1-10 GHz frequency range to the optical frequencies corresponding to 30-300 THz frequency range. The OSC uses undulators as a pickup and a kicker, and an optical amplifier for signal amplification, while the CEC uses an electron beam for all these functions.

Keywords: Accelerator Subsystems and Technologies, Beam dynamics, Beam Optics

---

[1] valebedev@jinr.ru

# 1. Introduction

In this paper we consider methods of particle cooling mostly concentrating on cooling of high energy heavy particles (protons or ions) in the high energy colliders. Further in all equations we assume protons – the most challenging case. To transit to ions, one needs to replace in the below equations the proton classical radius $r_p$ by $Z^2 r_p /A$, where $Z$ and $A$ are the charge and mass numbers of ion. Presently, there are two major methods of cooling: the electron cooling and the stochastic cooling. Up to recently, the stochastic cooling has been only operating at the microwave frequencies. Its cooling rates are limited by the width of its cooling band. A transition to much higher optical frequencies is expected to enable much faster cooling of dense colliding bunches. The stochastic cooling at the optical frequencies can be additionally subdivided into the optical stochastic cooling (OSC) and the coherent electron cooling (CEC). OSC and CEC are essentially extensions of microwave stochastic cooling operating in the 1-10 GHz frequency range to the optical frequencies corresponding to the 30-300 THz range. At these frequencies one cannot use usual electro-magnetic pickups and kickers. Instead, the OSC uses undulators for both the pickup and the kicker, and an optical amplifier for signal amplification; while the CEC uses an electron beam for all these functions.

The electron and stochastic cooling are based on completely different principles. The electron cooling is dissipative in its principle of operation and therefore the Liouville theorem is not applicable. That enables direct reduction of the beam phase space. The stochastic cooling is a "Hamiltonian" process which formally does not violate the Liouville theorem and cooling happens due to the phase space mapping so that phase space volumes containing particles are moved to the beam center while the rest mostly moves out. That makes stochastic colling rates strongly dependent on the beam particle density. As one will see each method has its own domain where it achieves a superior efficiency. The electron cooling is preferred at a smaller energy, and its efficiency weakly depends on the particle density in the cooled beam. The stochastic cooling is preferred at a higher energy, but its efficiency reduces fast with increase of particle density.

# 2. Low energy electron cooling

To understand principles of the electron cooling we start our consideration from a low energy electron cooling. The method of electron cooling was suggested by G. Budker [1]. It is based on a temperature exchange between a "hot" proton beam and a "cold" electron beam accompanying the proton beam in a straight section of storage ring. The first electron cooling demonstration was carried out in Novosibirsk, Russia at the beginning of seventies [2] and later implemented at a dozen of storage rings (see, *e.g.*, the review [3]).

### 2.1. Beam temperatures

The electron beam is formed in an electron gun and propagates in accompanying magnetic field to a collector. The magnetic field prevents the electron beam divergence due to its space charge fields and the corresponding effective temperature growth on the way from the gun to the collector. The magnetic field, if it is strong enough, also freezes out the transverse motion of electrons and basically switches off their hot transverse degrees of freedom from beam cooling. Because the longitudinal temperature is much smaller than the transverse one it results in very large increase of cooling force at small velocities while it relatively weakly affects cooling at large velocities.

We consider the case when the electron beam propagates in a longitudinal magnetic field of constant strength. Consequently, the transverse beam sizes stay constant. For an electron gun with good optics one can neglect a transverse velocity perturbation introduced by the gun. Then the transverse temperature is equal to the cathode temperature, $T_c$, while the longitudinal temperature (determined in the beam frame) is greatly reduced by the acceleration [4]:

$$T_\| \approx \frac{T_c^2}{2W} + 1.9 e^2 n_e^{1/3} . \tag{1}$$

Here $W$ is the energy of electrons, and $e$ and $n_e$ are their charge and density. The first term is related to the temperature of the cathode and is determined by the energy conservation in the absence of particle interaction. That results in that an energy difference of two electrons leaving the cathode does not change in the course of electrostatic acceleration in the lab frame, while the energy difference in the beam frame is decreasing inversely proportional to the beam energy (in non-relativistic case). The second term is related to the interaction of electrons. The beam acceleration in the Pierce gun is happening much faster than a plasma period [4]. Consequently, after acceleration the relative positions of electrons do not change much in the course of beam acceleration. Due to large temperature of electrons near the cathode their positions are almost random, and they stay random after the acceleration. That results in the correlation energy [5] which results in a temperature increase happening in about quarter of plasma period after acceleration. Typically, the resulting longitudinal temperature still stays much lower than the transverse one. Further thermalization between the horizontal and longitudinal degrees of freedom (Boersch effect) is much slower. In practical coolers this thermalization is suppressed by magnetic field and, typically, can be neglected. In most practical cases, with exception very low energy coolers, the second term in Eq. (1) is much higher than the first one. This term has a relatively weak dependence on the beam density and typically determines the energy spread in the electron beam. For beam density of $2 \cdot 10^8$ cm$^{-3}$ one obtains the longitudinal temperature of ~0.16 meV, while for typical thermo-cathode the transverse temperature is ~120 meV.

### 2.2. Cooling force

In the absence of magnetic field, the cooling force is described by a well-known expression [5]:

$$\mathbf{F}(\mathbf{v}) = \frac{4\pi n_e e^4 L_c}{m_e} \int f(\mathbf{v}') \frac{\mathbf{v}-\mathbf{v}'}{|\mathbf{v}-\mathbf{v}'|^3} d\mathbf{v}'^3 = \frac{4\pi n_e e^4 L_c}{m_e} \nabla_v \left( \int \frac{f(\mathbf{v}')}{|\mathbf{v}-\mathbf{v}'|} d\mathbf{v}'^3 \right) . \tag{2}$$

Here $\mathbf{v} = (v_x, v_y, v_z)$, is the proton velocity in the beam frame, $m_e$ is the electron mass, and $L_c$ is the Coulomb logarithm,

$$L_c = \ln\left(\frac{r_{max}}{r_{min}}\right), \quad r_{min} \approx \frac{2e^2}{m_e v_{eff}^2}, \quad r_{max} \approx \frac{v_{eff}}{\omega_p}, \quad v_{eff} = \max\left(|\mathbf{v}|, \bar{v}_e\right), \tag{3}$$

$\omega_p = \sqrt{4\pi n_e e^2 / m}$ is the electron plasma frequency, and $\bar{v}_e$ is the rms electron velocity. Eq. (2)

was obtained using the perturbation theory and is justified if $L_c \gg 1$. For a by-Gaussian distribution,

$$f(\mathbf{v}_\perp, \mathbf{v}_\parallel) = \frac{1}{(2\pi)^{3/2} \sigma_{v\parallel} \sigma_{v\perp}^2} \exp\left(-\frac{v_\parallel^2}{2\sigma_{v\parallel}^2} - \frac{v_\perp^2}{2\sigma_{v\perp}^2}\right), \tag{4}$$

the integral in Eq. (2) can be reduced to a single dimensional integral (so-called Binney formula, see, e.g., Ref. [6])):

$$\mathbf{F}(\mathbf{v}) = \frac{4\pi n_e e^4 L_c}{m_e} \nabla_v \left( \frac{2}{\sqrt{\pi}} \int_0^\infty \frac{\exp\left(-\frac{v_\parallel^2 t^2}{1+2\sigma_{v\parallel}^2 t^2} - \frac{v_\perp^2 t^2}{1+2\sigma_{v\perp}^2 t^2}\right)}{\sqrt{(1+2\sigma_{v\parallel}^2 t^2)(1+2\sigma_{v\perp}^2 t^2)^2}} dt \right), \tag{5}$$

where $\nabla_v$ denotes the gradient in the velocity space. Figure 1 presents results of numerical integration of Eq. (5) for $\sigma_{v\parallel} = \sigma_{v\perp}/20$. With further reduction of longitudinal temperature, the peak of longitudinal force is approaching the characteristic force:

$$F_r = \frac{4\pi n_e e^4 L_c}{m_e \sigma_{v\perp}^2}. \tag{6}$$

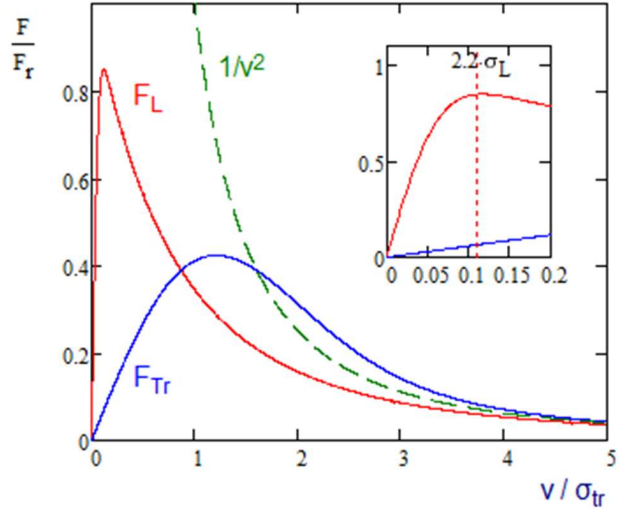

Figure 1: Dependence of longitudinal ($F_\parallel(v_\parallel, v_\perp=0)$ and transverse ($F_\perp(v_\parallel=0, v_\perp)$) cooling forces on particle velocity; $F_r = 4\pi n_e e^4 L_c / (m_e \sigma_{v\perp}^2)$, $\sigma_{v\parallel} = \sigma_{v\perp}/20$.

The maximum force is achieved at $v_\parallel / \sigma_{v\parallel} \simeq 2$. The peak of the transverse force of $\approx 0.45 F_r$ is achieved at $v_\perp \simeq 1.2\sigma_{v\perp}$. Sufficiently strong magnetic field magnetizes the transverse motion of electrons resulting in freezing out the transverse temperature [7]. If the longitudinal temperature is much smaller than the transverse one the magnetization yields very large increase of cooling force at small velocities. For the case when the transverse temperature is completely frozen out, the effective electron beam temperature is determined by Eq. (1). Then the maximum cooling force of $\sim e^2 n_e^{2/3}$ is achieved at the proton velocity of $\sim \sqrt{e^2 n_e^{1/3}/m_e}$. Note that at such small velocities the Coulomb logarithm $L_C \sim 1$ and the plasma perturbation theory is not applicable. The experimental studies [8, 9] verified correctness of such estimates.

The electron beam magnetization is very attractive for small energy machines. However, it is not helpful for hadron colliders where $\gamma \gg 1$ and the rms proton velocities are comparable to the rms transverse velocities in the electron beam. Thus, the magnetized cooling does not improve cooling of high amplitude particles, while, if present, it will overcool the proton beam core resulting in particle loss due to beam-beam effects. Therefore, we further consider only non-magnetized electron cooling.

### 3. High energy electron cooling

The highest energy electron cooling achieved to the present time is 4.3 MeV [10, 11]. It was used in the course of Tevatron Run II for cooling 8 GeV antiprotons. The electron beam was accelerated in an electrostatic accelerator. The maximum beam current was 500 mA. A typical beam current used for antiproton storage was about 100 - 200 mA with 6 mm electron beam radius. That supported beam cooling time of about 20 min. Using relativistic analog of Eq. (1) for beam current of 200 mA one obtains the relative momentum spread in the lab-frame of $9.2 \cdot 10^{-6}$. There is also the longitudinal momentum droop across beam radius driven by the electron beam space charge. For 200 mA beam current the potential difference between the beam center and its edge is 6 V. It drives the corresponding relative momentum difference of $1.2 \cdot 10^{-6}$. This is in striking difference to the lower energy coolers where the space charge driven momentum spread strongly exceeds the value set by the longitudinal temperature. Note that for the FNAL cooler the voltage ripple at the high voltage terminal represented the major source of longitudinal momentum spread. Its typical value [21] was about 150 V, rms, which corresponds to the rms relative momentum difference of $3.1 \cdot 10^{-5}$.

The next step in the proton energy increase requires the electron beam energy above tens of MeV. That is impossible to achieve with electrostatic acceleration. Consequently, two radically different approaches to the beam acceleration were suggested. The first approach suggests using the energy recovery superconducting linac [12], while the second approach uses an induction linac with electron beam injection into a ring where the electron beam circulates for many thousands turns [13]. That allows one to reduce the electron beam power to acceptable level. Ref. [14] considers another modification of ring based cooler where the electron beam is actively cooled by its synchrotron radiation produced in the high-field undulators. That greatly reduces the power of the linac required for acceleration of electrons to the energy of operation.

Before we consider these schemes in detail, we need to write down equations for the cooling rates of relativistic beams. The corresponding derivations were carried out in Ref. [13]. They assume Gaussian distribution for both the electron and proton beams, use Eq. (5) for computation of cooling force in the beam frame and account for a transition to a relativistic beam energy. In all practical cases the longitudinal velocity spread in the beam frame is much smaller than the transverse one. Then the corresponding longitudinal and transverse emittance cooling rates can be approximated as following[2]:

$$\lambda_\| \equiv \frac{1}{\theta_\|^2}\frac{d\theta_\|^2}{dt} \approx \frac{4\sqrt{2\pi}n_e r_e r_p L_c}{\gamma^4 \beta^4 \left(\Theta_\perp + 1.083\Theta_\| / \gamma\right)^{3/2} \sqrt{\Theta_\perp \Theta_\|}} L_{cs} f_0 ,$$

$$\lambda_\perp \equiv \frac{1}{\theta_\perp^2}\frac{d\theta_\perp^2}{dt} \approx \frac{\pi\sqrt{2\pi}n_e r_e r_p L_c}{\gamma^5 \beta^4 \Theta_\perp^2 \left(\Theta_\perp + \sqrt{2}\Theta_\| / \gamma\right)} L_{cs} f_0 ,$$

$$\frac{\Theta_\|}{\gamma\Theta_\perp} \leq 2 . \qquad (7)$$

Here

$$\Theta_\| = \sqrt{\theta_{\|e}^2 + \theta_{\|p}^2} ,$$
$$\Theta_\perp = \sqrt{\theta_{\perp e}^2 + \theta_{\perp p}^2} , \qquad (8)$$

are the effective longitudinal and transverse rms momentum spreads, $\theta_{\|p} \equiv \sqrt{\Delta p_\|^2}/p$ and $\theta_{\perp p} \equiv \sqrt{\Delta p_\perp^2}/p$ are the relative longitudinal and transverse rms momentum spreads in the proton beam, $\gamma$ and $\beta$ are the relativistic factors, $n_e = \gamma n_e'$ is the electron beam density in the lab frame, $r_e$ and $r_p$ are the classical electron and proton radii, $\theta_{\perp e} = \sigma_{v\perp}/(c\beta)$ and $\theta_{\|e} = \sigma_{v\perp}/(c\beta\gamma)$ are the relative rms longitudinal and transverse momentum spreads in the electron beam, respectively, $L_{cs}$ is the cooling section length, and $f_0$ is the revolution frequency of proton beam. These equations imply that the proton beam is completely inside the electron beam, and the horizontal and vertical rms transverse angles in the proton beam are equal. They also imply that if the electron beam radius, $r_{eb}$, is smaller than the maximum impact parameter of Eq. (3), $r_{eb}$ should be used instead of $r_{max}$.

Further, to make a simple best-case estimate, we assume that the proton beam is focused to the center of cooling section and has equal horizontal and vertical beta-functions; both transverse and longitudinal rms momentum spreads in the electron beam are smaller than in the proton beam; the cooling length is 2 times larger than the beta-functions of the proton beam in the center of cooling section: $L_{cs} = 2\beta_x^* = 2\beta_y^* \equiv 2\beta^*$; and the electron beam radius is two times larger than the rms proton beam size. The latter requirement is determined by a necessity to cool high amplitude particles. In relativistic colliders the longitudinal effective momentum spread is small and the second terms in parentheses of Eqs. (7) can be neglected. As result we obtain:

---

[2] In this document we are using the cooling rates for emittances and squared momentum deviation. The amplitude cooling rates are twice smaller than the emittance cooling rates.

$$\lambda_{\parallel} = \frac{2\sqrt{2} r_e r_p \beta^* L_c}{\sqrt{\pi} \gamma^2 \beta^2 \varepsilon_{pn}^2 \Theta_{\parallel p} C} \frac{I_e}{e},$$

$$\lambda_{\perp} = \sqrt{\frac{\pi}{2}} \frac{r_e r_p \beta^{*3/2} L_c}{\gamma^{5/2} \beta^{3/2} \varepsilon_{pn}^{5/2} C} \frac{I_e}{e}. \qquad \Theta_{\parallel} \ll \gamma \Theta_{\perp}. \qquad (9)$$

Here $C$ is the proton ring circumference, $\varepsilon_{pn}$ is the rms normalized proton emittance, $I_e$ is the electron beam current, and we assumed uniform density distribution across the electron beam. Assuming eRHIC parameters [15]: the proton beam beta-functions in the cooling section center of 60 m, the total cooling section length of 120 m, the proton beam energy of 275 GeV, the rms normalized emittance of 2.7 μm, ring circumference of 3.8 km, and the electron beam of 100 A one obtains the rms proton angles in the cooling section of 12 μrad and the transverse emittance cooling time of about 50 min. The longitudinal cooling strongly depends on the momentum spread in the proton beam. For the rms momentum spread of $10^{-3}$ one obtains the cooling time of about 20 minutes. Note that to achieve these cooling rates one needs the effective transverse angles in the electron beam significantly smaller than 10 μrad. That requires both small emittance of electron beam and very good straightness of its trajectory in the cooling area. Since cooling at small amplitudes is much faster the beam distribution does not stay Gaussian in the cooling process. The cooling leaves long non-Gaussian tails in the distribution. Note also that the magnetized cooling does not increase the average cooling rates if the rms electron angles are much smaller than the rms proton angles.

Any practical electron cooling for a collider has to avoid overcooling of small betatron and synchrotron amplitudes. Consequently, one needs to have the rms electron velocities close to the rms proton ones. That will result in a reduction of quoted above cooling rates by at least factor of 2.

The above obtained cooling rates need to be compared with the emittance growth rates due to intrabeam scattering (IBS) in the proton beam. For an ultra-relativistic beam, the IBS rates can be approximated as following (see Section 2.4.12 in Ref. [16]):

$$\frac{d}{dt}\begin{pmatrix} \varepsilon_x \\ \varepsilon_y \\ \sigma_p^2 \end{pmatrix} \approx \frac{N_p r_p^2 c}{4\sqrt{2}\beta^3 \gamma^3 \sigma_z} \left\langle \frac{\Psi(\theta_{xp}, \theta_{yp}) L_{cp}}{\sigma_x \sigma_y \sqrt{\theta_{xp}^2 + \theta_{yp}^2}} \begin{pmatrix} A_x \\ 0 \\ 1 \end{pmatrix} \right\rangle_s, \qquad (10)$$

where

$$\Psi(x,y) \approx 1 + \frac{\sqrt{2}}{\pi} \ln\left(\frac{x^2 + y^2}{2xy}\right) - 0.055\left(\frac{x^2 - y^2}{x^2 + y^2}\right)^2,$$

$N_p$ is the number of particles in a proton bunch, $\sigma_x$, $\sigma_y$, $\sigma_z$ are the rms bunch sizes in the horizontal, vertical and beam directions, respectively, $\theta_{xp}$ and $\theta_{yp}$ are the corresponding angular spreads, $L_{cp}$ is the Coulomb logarithm, $A_x = \left(D_x^2 + (\beta_x D_x' + \alpha_x D_x)^2\right)/\beta_x$ is the dispersion invariant, $\beta_x$ and $\alpha_x$ are the beta- and alpha-functions, and $D_x$, and $D'_x$ are the dispersion and its derivative. In the smooth lattice approximation one can easily obtain from Eq. (10) the emittance growth rates:

$$\begin{pmatrix} \lambda_x \\ \lambda_y \\ \lambda_p^2 \end{pmatrix} \equiv \begin{pmatrix} (1/\varepsilon_x) d\varepsilon_x/dt \\ (1/\varepsilon_y) d\varepsilon_y/dt \\ (1/\theta_{\|p}^2) d\theta_{\|p}^2/dt \end{pmatrix} \approx \frac{N_p r_p^2 c L_{cp}}{8\sqrt{\gamma} \sigma_z \varepsilon_{pn}^2} \begin{pmatrix} \sqrt{R_0/(\varepsilon_{pn} v_x^5)} \\ 0 \\ \left(\sqrt{v_x \varepsilon_{pn}/R_0}\right)/(\gamma \theta_{\|p}^2) \end{pmatrix}, \quad R_0 = \frac{C}{2\pi}, \quad (11)$$

where $v_x$ is the betatron tune. We also assumed $v_x \approx v_y$, $\beta_x \approx \beta_y \approx R_0/v_x$, $D_x \approx R_0/v_x^2$, $A_x \approx R_0/v_x^3$, $\sigma_x \approx \sigma_y \approx \sqrt{\varepsilon_{pn} \beta_x/\gamma}$, $\theta_{xp} \approx \theta_{yp} \approx \sqrt{\varepsilon_{pn}/(\beta_x \gamma)}$.

**Table 1: Main parameters of proton and electron beams**

| | |
|---|---|
| Proton beam energy, GeV | 275 |
| Proton ring circumference, m | 3834 |
| Number of particles in a proton bunch, $N_p$ | 6.9·10$^{10}$ |
| Number of proton bunches | 1160 |
| Cooling length section, $L_{cs}$, m | 120 |
| Normalized rms proton beam emittances, $\varepsilon_{pn}$ (x/y), μm | 2.7/0.47 |
| Proton beam rms momentum spread, $\theta_{\|p}$ | 7×10$^{-4}$ |
| Proton bunch rms bunch length, $\sigma_z$, cm | 6 |
| β-functions of proton beam at the cooling midpoint, $\beta^*$, m | 60 |
| IBS heating time for longitudinal emittance (Eq. (11) / Ref [15]), hour | 5 / 4 |
| IBS heating time for transverse emittance (Eq. (11) / Ref [15]), hour | 3.5/2 |
| Electron beam energy, MeV | 150 |
| Electron beam current, $I_e$, A | 100 |
| Ratio of electron beam mode emittances | 630 |
| Normalized electron beam rms mode emittances, $\varepsilon_{1n}/\varepsilon_{2n}$, μm | 68/0.108 |
| 4D-beta functions of electron beam in cooling section, m | 9.56 |
| Cathode diameter, mm | 11.5 |
| Cathode temperature, °C | 1050 |
| Longitudinal magnetic field in cooling section, G | 525 |
| Rms electron angles in cooling section, μrad | 6.2 |
| Rms electron beam size in cooling section, mm | 1.5 |
| Longitudinal emittance cooling time, hour | 0.3 |
| Transverse emittance cooling time, hour | 0.8 |

Comparing Eqs. (9) and (11) one can see that for fixed normalized emittances, tunes, electron beam current and particle number in a proton bunch the horizontal cooling rate decreases with energy as $\propto 1/\gamma^{5/2}$ while the corresponding IBS growth rate $1/\sqrt{\gamma}$. Thus, to compensate the IBS by electron cooling the electron beam current should grow proportionally to $\gamma^2$. Similarly, for the longitudinal degree of freedom the electron beam current should grow proportionally to $\sqrt{\gamma}$. The growth of reactive electron beam power has additional $\gamma$. Thus, the electron beam power and, consequently, difficulties of electron cooling grow as $\gamma^3$. Note that the transverse IBS growth rate can be additionally decreased by an increase of betatron tunes with energy. However, it results in higher IBS growth rate for the longitudinal degree of freedom.

The top part of Table 1 presents tentative major parameters of proton beam based on the eRHIC parameters of Ref. [15]. As one can see in the table a simple estimate of the IBS rates with Eq. (11) is quite close to the accurate calculations presented in Ref. [15]. The choice of electron beam parameters was driven by the above considered physics. It also accounted for the assumptions used in the derivation of Eq. (9).

There are two possibilities to focus the electron beam in the cooling straight. The first one is based on normal *xy*-uncoupled optics, while the second one is based on the longitudinal magnetic field accompanying the electron beam in the cooling section.

In the case of uncoupled electron beam focusing to determine an upper boundary of electron beam emittance we require that the rms transverse angles in the electron beam do not exceed the half of horizontal rms angle in the proton beam and the rms size would be twice larger than for the proton beam. That yields that the electron beam emittance should be equal to the proton beam emittance, and the electron beam beta-function should be 4 times larger than the proton beam beta-function in the cooling section center, i.e. 240 m. Further, we can assume that the electron beam focusing is supported by large number of quadrupoles or short solenoids, so that the focusing can be considered being smooth. Although the required value of the electron beam emittance looks reasonable, this choice does not look practical because the electron beam focusing is extremely weak. That makes the electron beam very vulnerable to perturbations and instabilities. In particular, the electric field of proton beam results in non-linear focusing of electron beam with characteristic distance of

$$L = \sqrt{\frac{\sqrt{2\pi}\beta^2\gamma^3\sigma_z\left(\varepsilon_{pn}\beta^*/\gamma\right)}{N_p r_e}} \approx 100 \text{ m}.$$

This focusing results in doubling transverse velocities in electron beam at distance of ~13 m thus eliminating any possibility of using this scheme in a real collider.

An electron beam transport in the accompanying magnetic field enables to resolve this problem [13]. To cancel a rotational kick appearing at the entrance of cooling solenoid the electron beam is magnetized at the gun cathode. That creates a vortex in the transverse beam particle distribution. This vortex is canceled when the beam enters a cooling solenoid. In the absence of magnetic field, the normalized beam emittance is set by the cathode temperature, $T_c$, and its radius $r_c$:

$$\varepsilon_n = \frac{r_c}{2}\sqrt{\frac{T_c}{m_e c^2}} \ . \tag{12}$$

Here we also imply that the gun optics is good enough so that it does not introduce additional emittance growth [13]. Magnetic field at the cathode couples the horizontal and vertical degrees of freedom. In this case the beam is described by the two mode emittances (see e.g. Section 2.2.5 in Ref. [16] or Ref. [17]) which values are set by the magnetic field at the cathode (see Appendix B in Ref. [17]):

$$\varepsilon_{1n,2n} = \frac{\varepsilon_n}{\sqrt{1+\Phi_r^2 \beta_0^2} \pm \Phi_r \beta_0} \ , \tag{13}$$

where $\beta_0 = a_e^2/(\varepsilon_n/\beta\gamma)$ is the effective beta-function, $a_e$ is the electron beam radius in the cooling section, $\beta$ and $\gamma$ are the relativistic factors, $B_0$ is the magnetic field in the cooling solenoid and $\Phi_r = eB_0/(2\gamma\beta m_e c^2)$ is its focusing strength. Here we also accounted that the magnetic field fluxes through the beam cross section are equal for the gun and the cooling section. That allowed us to express $\Phi_r$ through the beam size and magnetic field in the cooling section. The product of the mode emittances is equal to $\varepsilon_n^2$. The same as for uncoupled case a follow up to the above requirements yields that the electron beam emittance should be equal to the proton beam emittance, but the electron beam 4D-beta-function [17] is equal to:

$$\beta_e = \frac{4\beta_p^*}{\sqrt{R_\varepsilon}} \ . \tag{14}$$

Here $\beta_p^*$ is the proton beam beta-function in the cooling area, and

$$R = \frac{\varepsilon_{1n}}{\varepsilon_{2n}} = \frac{\sqrt{1+\Phi_r^2 \beta_0^2} + \Phi_r \beta_0}{\sqrt{1+\Phi_r^2 \beta_0^2} - \Phi_r \beta_0} \tag{15}$$

is the ratio of beam mode emittances. For $R = 630$ we obtain the electron beam beta-function $\beta_0$=9.56 m, and the mode emittances $\varepsilon_{1n}$=68 µm, $\varepsilon_{1n}$=0.108 µm. The required initial emittance, $\varepsilon_n$ = 2.7 µm, can be obtained from a general thermal cathode with temperature of 1050 C° and radius 11.5 mm. The required magnetic fields are: 525 G in the cooling section, and 9 G at the cathode.

In difference to the uncoupled focusing case the coupled focusing with accompanying magnetic field prevents the uncontrolled growth of transverse velocities in the electron beam. It can be almost completely cancelled if the cooling section length is equal to an integer number of Larmor periods. That enables to use such electron beam in a cooling ring where the beam comes through cooling section many thousand turns. In the above example an electron makes two Larmor revolutions at the cooling section length. The magnetic field in the cooling section can be increased, if required. However, such an increase will result in a larger ratio of the mode emittances resulting in more rigid requirements to the machine decoupling [13]. Note that the ring-based cooler considered in Ref. [14] has uncoupled focusing and very large beta-functions (150-300 m) in the cooling straight. Consequently, the considered above non-linear electron beam focusing by proton beam fields will result in a very fast growth of beam emittance with time, thus making such arrangement being impractical.

It is important to stress that the transverse velocities in the colliding proton bunches are large; so large that the transverse electron velocities in a beam produced in a well-designed electron gun can be smaller than the proton velocities. In this case the beam magnetization does not improve cooling of large amplitude protons, while may results in overcooling in the core thus worsening the beam-beam effects with subsequent increase of proton beam loss. Therefore, a magnetized electron cooling [7] should not be used for electron cooling in a high energy proton collider. Note that the value of considered above magnetic field is insufficient to make cooling magnetized.

As one can see from the above estimates the beam cooling required to counteract the intrabeam scattering in the electron-ion collider requires the peak beam current of about 50-100 A. The electron beam can be continuous as considered in Ref. [13] or bunched as suggested in Ref. [14]. There is no considerable margin in the cooling rates. Therefore, in the case of bunched beam the electron bunch length should be at minimum equal to the proton bunch length. For a 100 A peak current and 6 cm rms bunch length one obtains the electron bunch population of $3 \cdot 10^{11}$. This number is about an order of magnitude above what was obtained in the state-of the art electron SC linacs. Thus, presently, a usage of energy recovery linac for the electron beam cooling, as suggested in Ref. [12], looks extremely challenging.

Summarizing we can state that the high energy electron cooling looks as a possibility in the energy range required for the electron-ion collider. The most realistic scenarios are based on a ring-based cooler [13,14]. However, a construction of such cooler requires addressing extremely challenging problems ranging from achieving the straightness of magnetic field in the cooling section better than ~5 µrad at ~100 m length to the suppression of beam emittance growth and beam instabilities [13]. Presently, the electron cooling at energies above ~300 GeV looks extremely questionable.

## 4. Stochastic cooling of continuous beam

Stochastic cooling was suggested by Van der Meer [18]. It was the primary technology for cooling and accumulation of antiprotons in the SppS and Tevatron colliders. Up to recently the stochastic cooling was only demonstrated at the microwave frequencies, typically in the range 1 to 8 GHz. In this case, typically, the slip-factor of a storage ring is set so that to avoid an overlap of Schottky bands. It greatly simplifies tuning of cooling systems and removes unwanted coupling between different cooling systems. In this case, as will be seen below, the fastest cooling implies the choice of slip-factor, for which the Schottky bands are almost overlapped. For a fixed bandwidth further increase of band overlap does not yield significant increase of cooling rates [19] but may create additional problems with system tuning and operation.

First, we consider the general stochastic cooling theory for a continuous beam [20,21] which is equally applicable for cooling with and without band overlap. In the following sections we apply this theory to a bunched beam and consider corrections specific for the OSC and CEC which appear due to much higher frequency and technical restrictions of OSC and CEC.

Here we consider longitudinal cooling only. That allows one to understand major principles and limitations. Detailed description of the transverse cooling can be found in Refs. [20, 21]. There are three major methods of longitudinal stochastic cooling [22,23]. They are the Palmer cooling, the filter cooling and the transit-time cooling. Only the latter can be used at optical frequencies. Further we consider the Palmer cooling and the transit-time cooling, and do not consider the filter cooling because it cannot operate with band overlap.

### 4.1. Palmer cooling

An evolution of longitudinal distribution function, $\psi(x,t)$, during stochastic cooling of continuous beam is determined by the Fokker-Planck equation [20]:

$$\frac{\partial \psi(x,t)}{\partial t} + \frac{\partial}{\partial x}\left(F(x)\psi(x,t)\right) = \frac{1}{2}\frac{\partial}{\partial x}\left(D(x)\frac{\partial \psi(x,t)}{\partial x}\right) . \tag{16}$$

where $x \equiv \Delta p/p$ is the relative momentum deviation, $t$ is the time,

$$F(x) = \frac{1}{T_0}\sum_{n=-\infty}^{\infty}\frac{G(x,\omega_n(x))}{\varepsilon(\omega_n(x))}e^{2\pi i n \eta_{pk} x} \tag{17}$$

is the cooling force,

$$D(x,t) = \frac{N}{T_0}\sum_{n=-\infty}^{\infty}\frac{|G(x,\omega_n(x))|^2}{|\varepsilon(\omega_n(x,t))^2|}\sum_{k=-\infty}^{\infty}\frac{1}{|k\eta|}\psi\left(\frac{k-(1-\eta x)n}{k\eta},t\right) \tag{18}$$

is the diffusion,

$$\varepsilon(\omega,t) = 1 + N\int_{\delta \to 0_+}\frac{d\psi}{dx}\frac{G(x,\omega)e^{i\omega T_0 \eta_{pk} x}}{e^{i\omega T_0 (1+\eta x)}-(1-\delta)}dx \tag{19}$$

is the dielectric function, $T_0$ is the revolution time for the reference particle, $N$ is the number of particles, $G(x,\omega)$ is the generalized gain function, $\omega_n(x) = n\omega_0(1-\eta x)$, $\omega_0 = 2\pi/T_0$ is the revolution angular frequency, $\eta = \alpha - 1/\gamma^2$ is the ring slip-factor, $\alpha$ is the ring momentum compaction, $\eta_{pk}$ is the partial pickup-to-kicker slip-factor determined so that the change of pickup-to-kicker flight time relative to the reference particle is equal to: $\Delta T_{pk} = T_0 \eta_{pk} x$, and $\psi$ is normalized so that $\int \psi(x,t)dx = 1$. Here we neglected in Eq. (18) the contribution of amplifier noise to the diffusion, which is well justified for most practical systems. The sum over $k$ in Eq. (18) accounts for the band overlap. In the absence of the overlap only one term in the sum with $k = n$ is left, and the sum is equal to $\psi(x)/|n\eta|$. In the case of strong band overlap the Schottky noise over one band does not depend on the frequency. Consequently, in the cases of no band overlap and the strong band overlap Eq. (18) can be reduced to:

$$D(x) = \frac{N}{T_0} \begin{cases} \sum_{n=-\infty}^{\infty} \frac{|G(x,\omega_n(x))|^2}{|\varepsilon(\omega_n(x))|^2} \frac{\psi(x)}{|n\eta|}, & \text{no band overlap,} \\ \sum_{n=-\infty}^{\infty} |G(x,\omega_n(x))|^2, & \text{strong band overlap.} \end{cases} \quad (20)$$

Here in the bottom equation, we accounted that $\varepsilon(\omega) \approx 1$ which, as will be seen below, is well justified for a system operating with strong band overlap at the optimal gain. To make equations shorter, here and below we drop denoting of dependencies of functions on time.

Multiplying Eq. (16) by $x^2$ and integrating over $x$ one obtains an equation describing evolution of the rms momentum spread:

$$\frac{d\overline{x^2}}{dt} = -2\frac{G}{G_{ref}}\overline{F} + \left(\frac{G}{G_{ref}}\right)^2 \overline{D}, \quad (21)$$

where the integrals of cooling force and diffusion,

$$\overline{F} = -\int xF(x)\psi(x)dx,$$
$$\overline{D} = \int \psi(x)\frac{d}{dx}(xD(x))dx, \quad (22)$$

are calculated for an arbitrary reference gain $G_{ref}$. As one can see the first term in the right side of Eq. (21) is proportional to $G$ and the second one counteracts cooling and is proportional to $G^2$. It implies that there is optimal value of the gain where the cooling rate achieves its maximum. Equaling a derivative of Eq. (21) over $G$ to zero one obtains the optimal gain:

$$G_{opt} = G_{ref}\frac{\overline{F}}{\overline{D}}, \quad (23)$$

and the maximum rate of momentum cooling:

$$\left.\frac{d\overline{x^2}}{dt}\right|_{max} = -\frac{\overline{F}^2}{\overline{D}}. \quad (24)$$

Note that the maximum rate of momentum cooling does not depend on the choice of $G_{ref}$.

First, we consider the Palmer cooling. The gain function for the Palmer cooling can be presented in the following form: $G(x,\omega) = -xG'(\omega)$. In further calculations we assume the Gaussian distribution over momentum[3],

$$\psi(x) = \frac{1}{\sqrt{2\pi}\sigma_p}\exp\left(-\frac{x^2}{2\sigma_p^2}\right), \quad (25)$$

---

[3] Strictly speaking the distribution function does not stay Gaussian in the course of beam cooling. Although a usage of Gaussian distribution yields compact and clear results, an accuracy of such approximation depends on the details of cooling process and needs to be separately analysed.

and a perfectly phased rectangular band: $G'(\omega) = G'$ for $\omega \in [\omega_1, \omega_2]$ and zero otherwise; $\text{Im}(G') = 0$. We also set $\varepsilon(\omega) = 1$. Accuracy of such approximation will be discussed later.

Using Eq. (17) and replacing summation by integration one obtains the cooling force equal to:

$$F(x) = -\frac{G'}{\pi}(\omega_2 - \omega_1) x \Im(x),$$
$$\Im(x) = \frac{\sin(2\pi n_2 \eta_{pk} x) - \sin(2\pi n_1 \eta_{pk} x)}{2\pi (n_2 - n_1) \eta_{pk} x},$$
(26)

where $n_1 = \omega_1/\omega_0$ and $n_2 = \omega_2/\omega_0$ are the harmonic numbers at the boundaries of cooling band. The form-factor $\Im(x)$ is equal to 1 for small $x$ and reduces the cooling force for large momentum deviations. The momentum deviation, where the cooling force approaches zero the first time,

$$x_b = \frac{1}{2(n_2 + n_1)|\eta_{pk}|},$$
(27)

determines the cooling range. Outside of the cooling range $[-x_b, x_b]$ the cooling is replaced by heating. The partial slip-factor $\eta_{pk}$ has to be sufficiently small, so that cooling would be present in the required momentum aperture. Figure 2 shows behavior of cooling form-factor on the momentum for different ratios of upper and lower band boundaries.

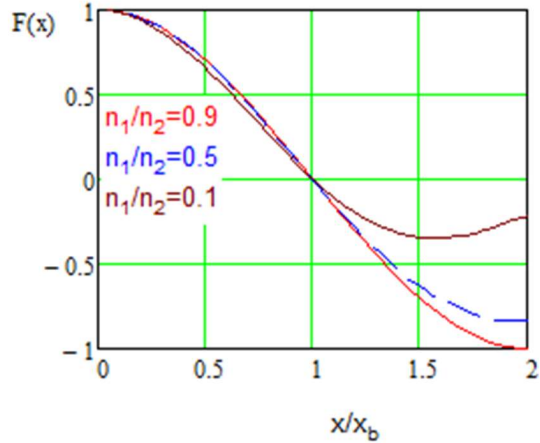

Figure 2: Dependence of cooling force form-factor $\Im(x)$ on particle momentum for Palmer cooling.

Good beam lifetime requires that the major fraction of beam particles is well within the cooling range. Therefore, in the calculation of cooling rate we can assume that $\Im(x)=1$ and, consequently, the cooling force is linear. Using Eq. (22) one finds the cooling force integral:

$$\overline{F} = 2G'(n_2 - n_1)\frac{\sigma_p^2}{T_0}.$$
(28)

Computation of $\overline{D}$ was carried out in Ref. [19]. For the cases of small and large band overlap the result is:

$$\bar{D} \approx \frac{3NG'^2}{T_0} \begin{cases} \dfrac{\sigma_p}{4\sqrt{\pi}\,|\eta|} \ln\left(\dfrac{n_2}{n_1}\right), & 2\eta\sigma_p n_2 \ll 1, \\ 2\sigma_p^2 (n_2 - n_1), & 2\eta\sigma_p n_1 \gg 1. \end{cases} \qquad (29)$$

Consequently, for the cooling rate at the optimal gain we obtain:

$$\lambda_{max} = -\frac{1}{\sigma_p^2} \left.\frac{d\overline{x^2}}{dt}\right|_{max} \approx \frac{2W}{3N} \begin{cases} \dfrac{8\sqrt{\pi}(n_2 - n_1)|\eta|\sigma_p}{\ln(n_2/n_1)}, & 2\eta\sigma_p n_2 \ll 1, \\ 1, & 2\eta\sigma_p n_1 \gg 1. \end{cases} \qquad (30)$$

where $W=(n_2-n_1)/T_0$ is the bandwidth of the cooling system.

Figure 3 presents the dependence of dimensionless cooling rate on the value of band overlap computed in Ref. [19] for systems with different bandwidths ($n_2/n_1$=1.1, 1.5, 2). As one can see the maximum cooling rate, $\lambda \approx 4W/(3N)$, is achieved at a minor band overlap, $\eta\sigma_p n_2 \approx 0.2$. Further increase of band overlap reduces the dimensionless cooling rate, $\bar{\lambda} = 3N\lambda/(2W)$ by about factor of 2.

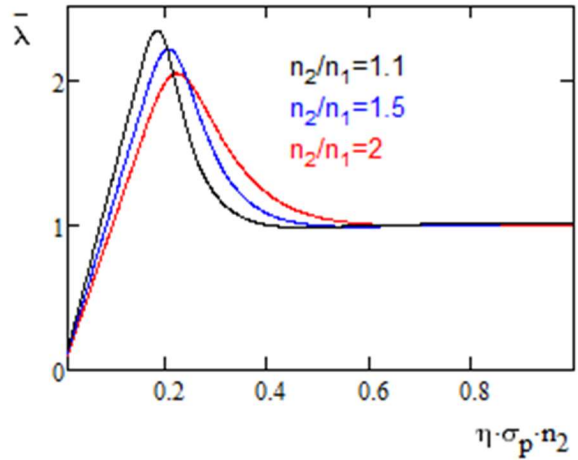

Figure 3: Dimensionless cooling rate, $\bar{\lambda} = 3N\lambda/(2W)$, for Palmer cooling at the optimal gain as a function of the band overlap at the upper band end for different bandwidths of the system.

In the above calculations we neglected that the dielectric function is different from 1. As it is pointed out in Ref. [21] it makes comparatively small correction. Such behavior is associated with the following. Both $\bar{F}^2$ and $\bar{D}$ are reduced by signal suppression in almost the same proportion: $\bar{F}^2 \propto \text{Re}(1/\varepsilon)^2$ and $\bar{D}^2 \propto 1/|\varepsilon|^2$. For a well-phased system, the difference in attenuations for both terms is comparatively small. The effect of signal suppression is more pronounced for systems operating far from the band overlap. However, as it was pointed out in Ref. [21] (see page 304), even in this case it makes less than 5% correction to the cooling rate for one-octave system operating at the optimal gain. In this case the maximum value of $|\varepsilon|$ being about 2 and, consequently, the strong signal suppression for both the cooling force and the diffusion. The effect of signal suppression can be much larger for a poorly phased cooling system. Finally, we note that for a system operating with strong band overlap the signal suppression is small and can be completely neglected.

### 4.2. Transit-time cooling

All standard microwave stochastic cooling methods are based on a subtraction of two signals: signals of two pickup plates in the Palmer or transverse cooling, and signals of two consecutive turns in the filter cooling. Consequently, the reference particle (or a particle which is completely cooled) does not create a signal on the kicker. That not only reduces the power of kicker amplifier but also reduces the diffusion introduced by cooling system, thus yielding faster cooling. However, in transition from the microwave frequencies to the optical frequencies we lose ability to create difference signals similar to the microwave cooling. In this case the transit-time cooling is the only practical choice. In the transit-time cooling the reference (already cooled) particle does not experience a kick because it comes at the right time when the kicker voltage excited by this particle is equal to zero. Particles which experience oscillations relative to the reference particle are coming and different times and therefore experience corrective kicks.

In the transit-time cooling the gain of the system should be purely imaginary: $G(x,\omega) = -iG(\omega)$. Then, for a system with rectangular band ($G(\omega) = iG_0$ for $\omega \in [\omega_1, \omega_2]$ and zero otherwise) one obtains from Eq. (17) the cooling force:

$$F(x) = \frac{2G_0}{T_0} \sum_{n=n_1}^{n_2} \text{Im}\left(e^{2\pi i n \eta_{pk} x}\right) = \frac{2G_0}{T_0} \frac{\sin\left(\pi \eta_{pk}(n_2 - n_1)x\right)}{\pi \eta_{pk} x} \sin\left(\pi \eta_{pk}(n_2 + n_1)x\right), \quad (31)$$

where we replaced summation by integration. The corresponding cooling range is:

$$x_b = \frac{1}{\eta_{pk}(n_2 + n_1)} \quad (32)$$

For small amplitude particles the cooling force is obtained by expending Eq. (31). It is equal to:

$$F(x) = \frac{2\pi \eta_{pk} G_0}{T_0}\left(n_2^2 - n_1^2\right)x . \quad (33)$$

Substituting it to Eq. (22) we obtain the force integral:

$$\overline{F} = \frac{2\pi \eta_{pk} G_0}{T_0}\left(n_2^2 - n_1^2\right)\sigma_p . \quad (34)$$

Similar to the previous section we find the diffusion,

$$D(x) = \frac{2NG_0^2}{T_0} \sum_{n=n_1}^{n_2} \sum_{k=-\infty}^{\infty} \frac{1}{|k\eta|} \psi\left(\frac{k-(1-\eta x)n}{k\eta}\right) . \quad (35)$$

Computation of $\overline{D}$ was carried out in Ref. [19]. For the cases of small and large band overlap the result is:

$$\bar{D} \approx \frac{2NG_0^2}{T_0} \begin{cases} \dfrac{\sigma_p}{4\sqrt{\pi}|\eta|\sigma_p} \ln\left(\dfrac{n_2}{n_1}\right), & 2\eta\sigma_p n_2 \ll 1, \\ (n_2 - n_1), & 2\eta\sigma_p n_1 \gg 1. \end{cases} \qquad (36)$$

Similarly to the above, we obtain the maximum cooling rate in the absence of band overlap and with strong band overlap (top and bottom equation, respectively):

$$\lambda_{max} \approx \frac{2\pi^2 \sigma_p^2}{x_b^2} \frac{W}{N} \begin{cases} \dfrac{4\sqrt{\pi}(n_2 - n_1)|\eta|\sigma_p}{\ln(n_2/n_1)}, & 2\eta\sigma_p n_2 \ll 1, \\ 1, & 2\eta\sigma_p n_1 \gg 1. \end{cases} \qquad (37)$$

Figure 4 presents the dependence of dimensionless cooling rate on the value of band overlap for systems with different bandwidths ($n_2/n_1$=1.1, 1.5, 2). As one can see, the same as for the Palmer cooling, the maximum cooling rate, $\lambda_{max} \approx (W/N)(5.3\sigma_p/x_b)^2$, is achieved at a minor band overlap, $\eta\sigma_p n \approx 0.25$.

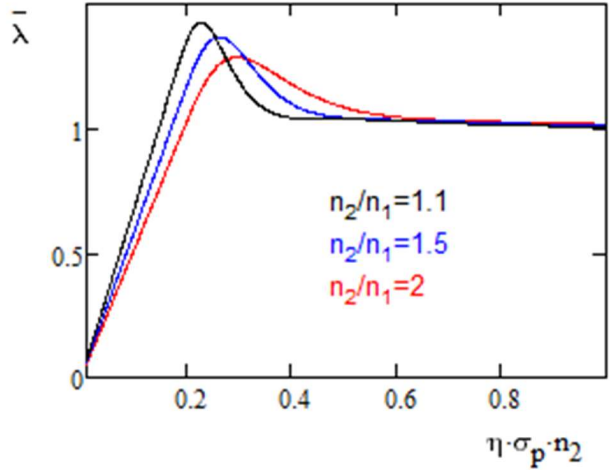

Figure 4: Dimensionless cooling rate, $\bar{\lambda} = Nx_{max}^2 \lambda / (2\pi^2 W \sigma_p^2)$, for transit-time cooling at optimal gain as a function of the band overlap at the upper band end for different bandwidths of the system.

As it should be expected, in difference to the Palmer cooling the cooling rate of transit-time cooling is additionally reduced by the ratio $\sigma_p^2/x_b^2$. This difference is associated with much larger diffusion for small amplitude particles in the transit-time cooling. As it was already stated the particle contribution to the diffusion is suppressed proportionally to its momentum deviation squared ($\propto x^2$) in the Palmer cooling but is not suppressed in the transit-time cooling.

In the case when the gain depends on a frequency and there is a strong band overlap, we can generalize the above equations. To achieve that we define the following quantities:

- the band central frequency –

$$\bar{f} = \int_0^\infty f \operatorname{Im}(G(f)) df \bigg/ \int_0^\infty \operatorname{Im}(G(f)) df , \qquad (38)$$

- the cooling range –

$$x_b = f_0 / (2\eta_{pk} \bar{f}) , \qquad (39)$$

- and the effective bandwidth of the system –

$$W_{eff} = \left( \int_0^\infty \operatorname{Im}(G(f)) df \right)^2 \bigg/ \int_0^\infty |G(f)|^2 df , \qquad (40)$$

where $f_0 = 1/T_0$ is the revolution frequency. Performing simple calculations similar to the above for the rectangular band one obtains the equation analogues to the bottom equation in (37):

$$\lambda_{max} \approx \frac{2\pi^2 \sigma_p^2}{x_b^2} \frac{W_{eff}}{N} , \quad 2\eta\sigma_p \frac{\bar{f}}{f_0} \gg 1 . \qquad (41)$$

For the rectangular band these quantities coincide with corresponding quantities introduced earlier.

We note that in the case of transit-time cooling with strong band overlap the diffusion does not depend on the momentum deviation, $x$. Therefore, initially Gaussian distribution stays Gaussian for the entire time of cooling process.

## 5. Stochastic cooling of bunched beam

Now let us consider how the optimal gain and the maximum cooling rate are changed for a bunched beam cooling. In the case of ideally linear synchrotron motions the beam spectrum consists of infinitely narrow lines at the combinations of revolution and synchrotron frequencies, $\sum_{n,m} \delta(\omega - \omega_0(n + mv_s))$. In this case the stochastic cooling is impossible due to an absence of good missing [23]. However, in real world small non-linearities and intrabeam scattering result in an overlap of different synchrotron sidebands due to very large values of $m$ present at very high frequencies where stochastic cooling works. The non-linearities come from non-linearity of RF focusing and electrical field excited by the beam in the vacuum chamber. Additional support comes from the intrabeam scattering, which breaks the particle synchrotron motion phase, resulting in additional smearing of high order synchrotron lines. Therefore, in practical conditions of microwave stochastic cooling the bunch spectrum at the cooling frequencies is continuous, and the theory developed for the continuous beam can be applied to the bunched beam cooling with minimal adjustments.

Similar to the case of continuous beam we can neglect $\varepsilon$. Then the cooling force and its integral do not depend on particle local density and, consequently, Eqs. (26), (28), (33) and (34) stay valid. The diffusion coefficient is proportional to the local particle density. Replacing the longitudinal particle density of continuous beam, $N/T_0$, in Eq. (35) by the longitudinal density of bunched beam with Gaussian longitudinal distribution,

$$\left(N/(\sqrt{2\pi}\sigma_t)\right)\exp(-t^2/2\sigma_t^2)\ ,$$

and performing averaging over bunch length we obtain the diffusion integral for the bunched beam:

$$\bar{D}_b = \bar{D}\int_{-\infty}^{\infty}\frac{T_0 e^{-t^2/2\sigma_t^2}}{\sqrt{2\pi}\sigma_t}\frac{e^{-t^2/2\sigma_t^2}}{\sqrt{2\pi}\sigma_t}dt = \frac{T_0}{2\sqrt{\pi}\sigma_t}\bar{D}\ . \tag{42}$$

Here $\sigma_t$ is the rms bunch duration, and $\bar{D}$ is the diffusion integral computed for a continuous beam with the same number of particles.

To obtain the cooling rate for a bunched beam one has to take into account that in the linear RF the synchrotron motion reduces both the single particle cooling rate and the diffusion by two times. The single particle cooling rate is reduced because the average momentum deviation squared is reduced by two times due to synchrotron motion. The effect of diffusion is reduced because the beam heating introduced by diffusion is equally divided between two degrees of freedom of oscillatory synchrotron motion (or, in other words, between potential and kinetic energies). We also need to note that for a Gaussian distribution the shape of momentum distribution does not depend on the longitudinal coordinate inside the bunch. Therefore, it is decoupled from the longitudinal coordinate, and should not be additionally accounted in averaging along the bunch. Thus, to obtain the cooling rate of bunched beam one has to multiply the cooling rates of continuous beam presented in Eqs. (30) and (37) by factor

$$F_{bb} = \sqrt{\pi}\sigma_t/T_0\ . \tag{43}$$

For the proton beam parameters of Table 1 and 4 to 8 GHz bandwidth we obtain that even in the most optimal case of band overlap, $\eta\sigma_p n_2 = 0.3$ the cooling time is $\approx$130 hour. It will also require an increase of $\eta\sigma_p$ by about 3 times. Thus, the microwave stochastic cooling of proton beam in a collider delivers cooling rates more than an order of magnitude below the needed value. Consequently, if any kind of stochastic cooling is going to be used its bandwidth has to be more than 0.5 THz.

## 6. Optical stochastic cooling

With transition from the microwave stochastic cooling [22,23] to the Optical Stochastic Cooling (OSC) [24] or to the Coherent Electron Cooling (CEC) [25 - 27], which operate at orders of magnitude higher frequencies, a cooling system operation without band overlap becomes impossible. That excludes the filter cooling. Technological difficulties (like an absence of electronic circuits for signal summation and subtraction) make the Palmer cooling cumbersome. Thus, the transit-time cooling is left as the only practical choice.

In the OSC a particle radiates electromagnetic wave in the first (pickup) undulator. Then, this wave is amplified in optical amplifier (OA) and refocused to the second (kicker) undulator as shown in Figure 5. The particle beam is separated from the radiation by four-dipole chicane creating space for optical amplifier and optical lenses. The chicane introduces a delay equal to a delay of radiation in the optical system, so that a particle would interact with its own radiation amplified by optical amplifier. In further consideration we assume that the chicane is in the horizontal plan and there are no *x-y* coupling terms in the chicane.

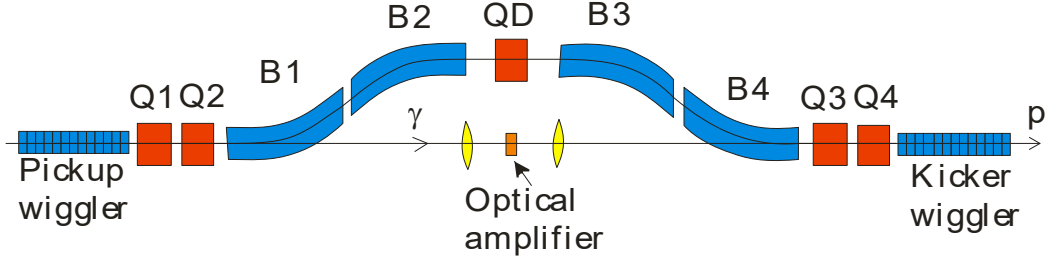

Figure 5: Layout of the OSC system.

We also assume that the optical system compensates for the depth of field [28] and that the bandwidth of the amplifier is sufficiently wide. So that the system bandwidth would be determined by number of undulator periods, $\Delta f/f \approx 1/n_u$, where $n_u$ is the number of undulator periods. In this case a particle, which is shifted longitudinally by distance $s = \psi/k_0$ from the reference particle[4] in the kicker undulator, experiences the longitudinal kick equal to:

$$\Delta p = \frac{e\beta c K_u E_0}{\gamma} \mathrm{Re}\left(\int \exp\left(i\left(\omega_0 t - k_0 z(t) - \psi\right)\right)\sin\left(k_u \beta c t\right) dt\right)$$

$$= \frac{e E_0 K_u L_u}{4\pi n_u c \gamma} \begin{cases} (2\pi n_u - |\psi|)\sin\psi, & |\psi| \leq 2\pi n_u \\ 0, & |\psi| \geq 2\pi n_u \end{cases}$$

(44)

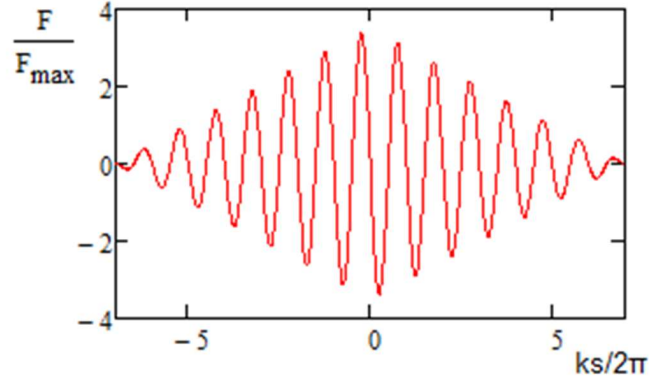

Figure 6: Dependence of cooling force on the longitudinal particle position relative to the reference particle.

Here $k_0$ is the wave number of forward radiation, $\psi = k_0 s$ is the particle phase relative to the wave, $E_0$ is the amplitude of e.-m. wave generated by a particle in the pickup undulator, amplified in the OA and focused into the kicker undulator, $\omega_0 = k_0 c$ is the wave angular frequency, $\gamma$ and $\beta$ are the particle relativistic factors, $K_u$ is the undulator parameter, $n_u$ and $L_u = 2\pi n_u / k_u$ are the number of undulator periods and its total length,

$$k_u = \frac{k_0}{2\gamma^2}\left(1+\frac{K_u^2}{2}\right)$$

is the wave number of undulator, and the integration in Eq. (44) is performed over the kicker undulator length. To simplify integration in Eq. (44) we assumed that $K_u \ll 1$. That allowed us to neglect the longitudinal particle displacement associated with the transverse particle motion in the magnetic field of kicker undulator. However, as more accurate consideration shows, that does not limit the generality of the results presented below. As an example, Figure 6 shows a plot of cooling force for the 7-period wiggler. Note that the synchrotron radiation frequency is decreased with an increase of angle between the radiation direction and the forward direction. Consequently, there are two bandwidths which characterize the OSC system. The first one, $\Delta f/f \approx 1/n_w$, is determined by the number of undulator periods, and the second one is determined by subtending angle of the radiation and the bandwidth of the optical amplifier. As it was stated above, we assume that the former is smaller and therefore it determines the cooling bandwidth introduced in the above-presented theory of stochastic cooling. Although the radiation frequency is decreasing with radiation angle the focusing of radiation to the kicker undulator yields that the moving particle sees resulting e.-m. wave having the same period independently from the direction of incoming wave.

---

[4] The reference particle has zero betatron and synchrotron amplitudes. It does not experience a cooling kick in the kicker undulator and is phased to maximize cooling force for nearby particles in the space of velocities.

For computation of cooling rate of small betatron and synchrotron oscillations one can leave only linear term in the Eq. (44) expansion. That results in:

$$\frac{\delta p}{p} = -\kappa k_0 s, \quad \kappa = \frac{eE_0 K_u L_u}{2pc\gamma}. \tag{45}$$

The particle longitudinal displacement depends on its momentum and betatron deviations. In the linear approximation one can write:

$$s = M_{51}x + M_{52}\theta_x + \left(M_{56} - \frac{L_{pk}}{\gamma^2}\right)\frac{\Delta p}{p}, \tag{46}$$

where $M_{5n}$ are the elements of 6x6 transfer matrix from pickup to kicker, $L_{pk}$ is the pickup-to-kicker distance, x, $\theta_x$ and $\Delta p/p$ are the particle coordinate, angle and relative momentum deviation in the pickup undulator. In the absence of betatron oscillations the particle transverse position and angle are determined by dispersion in the pickup undulator. Consequently, one obtains:

$$s = S_{pk}\frac{\Delta p}{p}, \quad S_{pk} = M_{51}D + M_{52}D' + M_{56} - \frac{L_{pk}}{\gamma^2}, \tag{47}$$

where $D$ and $D'$ are the horizontal dispersion and its derivative in the pickup undulator, and we assume that the vertical dispersion and its derivative are equal to zero. Combining Eqs. (45) and (47) one obtains the emittance cooling rate for the longitudinal degree of freedom in the linear approximation [28]:

$$\lambda_s = f_0 \kappa k_0 S_{pk} \ . \tag{48}$$

Here $f_0$ is the revolution frequency. In the absence of coupling between the horizontal and longitudinal degrees of freedom ($D = D' = 0$) the transverse cooling rates are equal to zero. The rate-sum theorem (see Section 2.2.5 in Ref. [16]) states that the sum of cooling rates is not changed if cooling rates are redistributed between degrees of freedom. Consequently, the sum of rates can be found from Eq. (48) if one sets $D = D' = 0$. That yields the sum[5]:

$$\lambda_s + \lambda_x + \lambda_y = f_0 \kappa k_0 \left( M_{56} - \frac{L_{pk}}{\gamma^2} \right) \ . \tag{49}$$

A direct proof of this statement in the application to the OSC can also be found in Ref. [28]. Considering that the sum of cooling rates is not changed one obtains the sum of transverse emittance cooling rates:

$$\lambda_1 + \lambda_2 = -f_0 k_0 \kappa \left( D M_{51} + D' M_{5,2} \right) = f_0 k_0 \kappa \left( M_{56} - S_{pk} - \frac{L_{pk}}{\gamma^2} \right) \ . \tag{50}$$

Typically, the term $L_{pk}/\gamma^2$ is small and we neglect it in further consideration. Then Eqs. (48) and (49) yield the ratio of cooling rates:

$$\frac{\lambda_1 + \lambda_2}{\lambda_s} = \frac{M_{56}}{S_{pk}} - 1 \tag{51}$$

The value of $M_{56}$ is determined by the cooling chicane optics and does not depend on the rest of the ring. Thus, for a given chicane and its $M_{56}$ the ratio is uniquely determined by the dispersion and its derivative in the pickup undulator (see Eq. (47)). Further we assume that the ring operates on a coupling resonance and has sufficiently strong coupling terms. In this case the generalized horizontal beta-functions [17] of two modes are equal, $\beta_{1x} = \beta_{2x}$. Then the cooling rates of modes 1 and 2 are equal. An example of such lattice is presented in Ref. [29].

As shown in Ref. [28] only particles with $k_0 s \leq \mu_0 \approx 2.405$ are cooled to the center of the distribution. Here $\mu_0$ is the first root of zeroth order Bessel function, $J_0(x)$. Accounting Eq. (47) we obtain the maximum momentum deviation which will be cooled – the so called the cooling range:

$$x_b \equiv \left( \frac{\Delta p}{p} \right)_{max} = \frac{\mu_0}{k_0 S_{pk}} \ . \tag{52}$$

Eqs. (48) and (50) do not account effect of other particles which introduce diffusion and limit the cooling as it was discussed in Section 4.2. The calculations are straightforward and we refer a reader to Appendix of Ref. [29] for details. Here we present the final result for the maximum achievable cooling rate of longitudinal motion:

---

[5] We assume that the ring lattice can be strongly coupled. In this case instead of horizontal and vertical cooling rates one must use cooling rates of modes 1 and 2.

$$\lambda_{s\_max} = \frac{3\mu_{01}{}^2 k_0 \sigma_s}{2\sqrt{\pi} n_u N n_{\sigma s}{}^2} , \qquad (53)$$

where $n_{\sigma s} = x_b / \sigma_p$ is the dimensionless longitudinal cooling range[6]. This equation can be also obtained from Eqs. (37) and (43). As one can see the cooling rate decreases quadratically with an increase of the cooling range $n_{\sigma s}$. The same equations are justified for the transverse cooling with the longitudinal cooling range being replaced by corresponding cooling ranges for the transverse degrees of freedom. Important to note that details of cooling in other planes do not affect the maximum cooling rate expressed by Eq. (53). Insufficient values of the cooling ranges may negatively affect the beam lifetime. For practical conditions of electron-proton collider the longitudinal cooling range should be above ~3-4, and transverse cooling ranges should be above ~4-5 to mitigate the beam-beam effects.

Table 2: Tentative parameters of optical stochastic cooling

| | |
|---|---|
| Wavelength of forward radiation, $2\pi/k_0$, μm | 4.79 |
| Number of undulator periods, $n_u$ | 25 |
| Peak undulator magnetic field, $B_0$, T | 12 |
| Length of undulator period, $L_u$, m | 0.75 |
| Undulator parameter, $K_u = eB_0/(mc^2 k_u)$ | 0.458 |
| Chicane $M_{56}$, mm | 1.3 |
| Dimensionless cooling range, $n_{\sigma s} \equiv x_b / \sigma_p$ | 6 |
| Energy loss due to synchrotron radiation in one undulator, meV | 11.4 |
| Gain in optical amplifier, dB | 40 |
| Average power of optical amplifier, W | 12.3 |
| Peak power of optical amplifier | 382 W |
| Emittance cooling times, $\lambda_1 / \lambda_2 / \lambda_s$, hour | 1.1/1.1/1.1 |
| Half-size of synchrotron radiation spot in the kicker undulator, $\rho_x/\rho_y$, mm | 2.1/1.5 |
| Emittance cooling time at the optimal gain, hour | 0.6 |

Table 2 presents the tentative parameters of the OSC system for an electron-ion collider with proton beam parameters presented in Table 1. As one can see the proposed system operates below the optimal gain. That leaves some margin for imperfections unavoidably resulting in a loss of cooling efficiency in a real system. The most challenging problem of the proposed system is related to an optical amplifier which has to have a small signal delay the and large gain (~$10^4$ in power). Another significant problem with OSC is a small energy range where a given system can operate. The wavelength is proportional to $1/\gamma^2$. Consequently, a change of beam energy requires either replacement of optical amplifier or replacement of undulator. The OSC operates well only for highly relativistic particles. The effectiveness of OSC drops fast with energy decrease making it impractical below ~ 200 GeV. Contrarily, an increase of beam energy would significantly simplify the optical amplifier (and undulator if required), so that above ~1 TeV a passive OSC could be sufficient [29]. We also need to note that the gain of optical amplifier has to cover the entire length of proton bunch ($\Delta t \geq 1$ ns). That makes a usage of optical parametric amplifiers challenging.

---

[6] This definition of cooling range is somewhat different from definition of Eq. (39) which determines the cooling boundary where the cooling force changes sign; while this definition considers the cooling range as a boundary where cooling rate changes its sign. The former was introduced for a continuous beam and represents a correct choice there; while the latter was introduced for the bunched beam.

Passive OSC was recently demonstrated with electrons [30]. This experiment clearly demonstrated the power of the method.

## 7. Coherent electron cooling

The coherent electron cooling (CEC) [25] was suggested to address the fast decrease of electron cooling force with an increase of proton velocity spread. First attempts to find a practical scheme were aimed at relatively small energy. They did not deliver a practical scheme for quite long time. The breakthrough happened at the end of 2000's with transition to relativistic energies and a suggestion to use the FEL (free electron laser) as an amplifier [26]. That addressed the main problem of CEC how to switch on and switch off the amplification (instability) in controllable fashion. However, after careful examination it was understood that the narrow band of FEL (~0.5%) and a quite short length of the electron bunch (~1/100 of proton bunch length) limit cooling rates to approximately the same level as it was already achieved in the bunched beam microwave stochastic cooling at RHIC [32]. Further investigations of CEC revealed two other schemes which are expected to have much wider bandwidth (up to ~50%). They are the micro-bunched electron cooling [27] and the cooling based on the plasma-cascade instability [33].

Same as OSC both cooling schemes operate at the same principle as the stochastic cooling and therefore can be described within the same theoretical framework. Moreover, similar to the OSC the CEC is based on the longitudinal kicks. Consequently, the transverse cooling is achieved by coupling transverse and longitudinal degrees of freedom. It is achieved by introduction of non-zero dispersion in the "pickup" and the "kicker" as described in the OSC section above.

**Table 3: Tentative parameters of electron beam for CEC**

| | |
|---|---|
| Number of electrons per bunch, $N_e$ | 6.3·10⁹ |
| Rms length of electron bunch, $\sigma_{ze}$ | 4 mm |
| Electron rms transverse beam size, $\sigma_{tr} = \sigma_x = \sigma_y$, mm | 0.6 mm |
| Dynamic gain, $G_d$ | 25 |
| Rms gain length, $\sigma_g$ | 4 mm |
| Kicker section length | 40 m |
| Phase advance of plasma oscillations on the kicker length | 30 deg |

All existing proposals of CEC are based on superconducting energy recovery linacs [34] which can deliver required transverse emittances and momentum spreads, but cannot deliver required number of particles in the bunch. To create a desired amplification, one need to have sufficiently large peak current. As result the electron bunch length is much shorter than the proton bunch length. That reduces the cooling rates in proportion of bunch length ratio. In further estimates we assume the electron bunch parameters of Ref. [35] presented in Table 3. The proton bunch parameters are the same as in Table 1. Here we will not consider each CEC scheme in detail, but rather consider phenomena which are common for all presently considered schemes and which represent major limitations on the cooling rates.

In our consideration we assume the most optimistic scenario where only two key limitations play a role. They are (1) the cooling rate reduction due to short length of electron bunch and (2) the saturation of the amplifier. Our consideration will be based on a single dimensional model which is well justified for the FEL based cooling, but can be considered only as a zero approximation in the case of micro-bunched cooling and plasma-cascade cooling where due to much larger bandwidth 3D analysis is required. Presently a development of such 3D models is at the initial stage, and a further study may yield additional limitations.

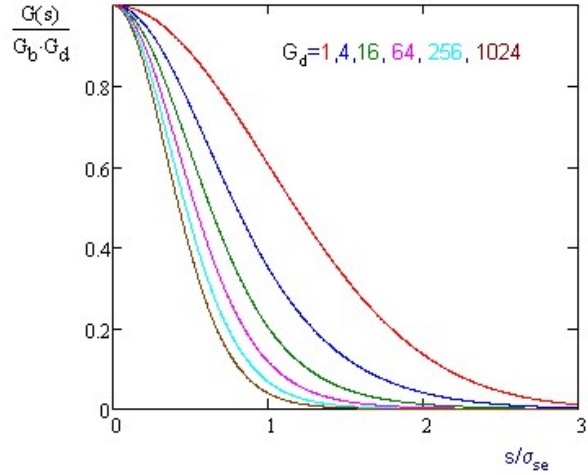

Figure 7: Variation of gain along the electron bunch for different values of dynamic gain, $G_d$=1,4,16,64,256,1024.

Before we move to the computation of how the electron bunch length affects the cooling, we consider how the gain is distributed along the electron bunch. We assume Gaussian distribution of the particle density with rms bunch length $\sigma_{se}$. In the first approximation, the total gain of CEC can be split into ballistic gain, $G_b$, which value is not related to amplification of plasma oscillations and is proportional to the bunch longitudinal density, and the dynamic gain, $G_d$, which value grows exponentially with time of instability development, $dG_d/dt = \lambda G_d$. Solving this differential equation, substituting the boundary conditions and combining the results one obtains the dependence of gain on the longitudinal coordinate along an amplifier:

$$G(s) \approx G_b e^{-s^2/2\sigma_{se}^2} \exp\left(\ln(G_d) e^{-s^2/2\sigma_{se}^2}\right), \quad (54)$$

where $G_b$ and $G_d$ are the gains in the bunch center. Figure 7 shows the gain variation along the Gaussian electron bunch for different dynamic gains. The gain length logarithmically depends on

$G_d$. The half-width half-maximum width can be approximated by the following equation:

$$\Delta s_{1/2} \approx \sigma_{se}\sqrt{\frac{2\ln(2)}{1+\ln(G_d)}} \ . \tag{55}$$

For $G_d = 1$ the gain distribution coincides with the bunch density distribution, and becomes twice narrower for $G_d = 25$. We will use $G_d = 1$ in further estimates. That is expected to be a good approximation for the microbunched cooling where the cooling force depends linearly on particle density. The cooling based on plasma cascade instability is expected to be very sensitive to the longitudinal density, and, consequently, one should expect $G_d \gg 1$. Note that the gain narrowing will be much more pronounced if the density distribution has narrow peaks. By design the CEC based on the plasma cascade instability has larger dynamic gain than the micro-bunch CEC, and, consequently, should experience larger shortening of the gain length.

Now we find how the electron bunch length, which is much shorter than the proton bunch length, affects the cooling rate. We assume that the gain along the electron bunch is described by Gaussian distribution with rms length $\sigma_g$, and the core particles are in the linear part of cooling force. Consequently, the cooling force along the proton bunch can be presented in the following form:

$$F(x,s) = \frac{dF}{dx} e^{-s^2/2\sigma_g^2} x, \quad \frac{dF}{dx} = \frac{4\pi\eta_{pk}}{T_0}\sum_{n=1}^{\infty}\text{Im}(G_n) \tag{56}$$

where $G_n$ is the gain in the center of the electron bunch. Similarly for the diffusion we have:

$$D(x,s) = \frac{C}{\sqrt{2\pi}\sigma_{sp}} D_0 e^{-s^2/\sigma_g^2} e^{-s^2/2\sigma_{sp}^2}, \quad D_0 = \frac{2N_p}{T_0}\sum_{n=1}^{\infty}|G_n|^2 \ . \tag{57}$$

Here $D_0$ is the diffusion for the continuous electron beam, and we assume that the electron bunch is in the center of the proton bunch which creates maximum average cooling. As one can see the diffusion in the proton bunch center is amplified by factor $C/(\sqrt{2\pi}\sigma_{sp})$ due to longitudinal density increase and is decreasing proportionally to the proton bunch density; and its value is proportional to square of the gain, i.e. $\exp(-s^2/\sigma_g^2)$.

In further estimate we neglect that the bunch does not stay Gaussian in the cooling process and compute the overall cooling rate for initially Gaussian beam. Similarly, to Eq. (24) we obtain the overall momentum cooling rate:

$$\left.\frac{d\overline{x^2}}{dt}\right|_{\max} = -\frac{1}{2}\frac{\overline{F^2}}{\overline{D}}, \tag{58}$$

where factor 1/2 appears due to linear longitudinal focusing which results in redistribution of heating and cooling powers between kinetic and potential energies, and $\overline{F}$ and $\overline{D}$ additionally account for averaging along the longitudinal coordinate:

$$\bar{F} = \int \frac{e^{-s^2/2\sigma_{sp}^2}}{\sqrt{2\pi}\sigma_{sp}} \frac{dF}{dx} e^{-s^2/2\sigma_g^2} x\psi(x) dx ds = \frac{\sigma_g}{\sqrt{\sigma_{sp}^2 + \sigma_g^2}} \frac{dF}{dx} \sigma_p^2,$$

$$\bar{D} = \int \frac{e^{-s^2/2\sigma_{sp}^2}}{\sqrt{2\pi}\sigma_{sp}} \frac{D_0 C e^{-s^2/\sigma_{sp}^2} e^{-s^2/\sigma_g^2}}{\sqrt{2\pi}\sigma_{sp}} ds = \frac{\sigma_g}{\sqrt{\sigma_{sp}^2 + \sigma_g^2}} \frac{CD_0}{2\sqrt{\pi}\sigma_{sp}}.$$

(59)

Substituting these equations into (58) and using Eqs. (56), (57) and definitions (38) - (40) we obtain the longitudinal emittance cooling rate:

$$\lambda_{max} \approx \frac{2\pi^2 \sigma_p^2}{x_b^2} \frac{W_{eff}}{N} \frac{\sqrt{\pi}\sigma_{sp}}{C} \frac{\sigma_g}{\sqrt{\sigma_g^2 + \sigma_{sp}^2}}, \quad 2\eta\sigma_p \frac{\bar{f}}{f_0} \gg 1.$$

(60)

As one can see for $\sigma_g \ll \sigma_{sp}$ the cooling rate is additionally reduced by $\sigma_{sp}/\sigma_g$ times. We need to note that the cooling rate in the proton bunch center is faster by $\approx \sigma_{sp}/\sigma_g$ times than its average. That results in overcooling of small synchrotron amplitudes and makes the longitudinal beam distribution peaked in the bunch center. This effect may worsen the beam lifetime since the diffusion due to intrabeam scattering has weak dependence on the synchrotron amplitude. The problem can be mitigated by sweeping the electron bunch along the proton bunch, but it will result in a reduction of average cooling rate. Eq. (60) also does not account for noise in the electron beam.

The peak current densities in electron and proton beams are quite close. In the case of noninteracting electrons this would double the diffusion and reduced the cooling rate by factor of 2. However, for considered momentum spreads in the electron beam [36,37] there is very strong particle interaction through the beam space charge which in the thermal equilibrium strongly suppresses the shot noise. Consequently, one could expect smaller contribution of electrons into the diffusion. However, together with suppression of shot noise comes disappearance of Landau damping. Consequently, the beam becomes very susceptible to instabilities and details of its generation, acceleration and transport. That may strongly increase the diffusion. Presently, there is no convincing measurements or analysis which would support a reduction or increase of electron beam noise below or above the shot noise. Thus Eq. (60) should to be considered describing an optimistic case.

Using the cooling force presented in Ref. [35] for the microbunched cooling and the beam parameters of Table 3 one obtains: the effective bandwidth - $W_{eff}$=74 THz, the average frequency - $f_a$=47 THz and the cooling range size - $s_b$=3.2 μm. The cooling range, $n_\sigma = s_b/(S_{pk}\sigma_p)$, determines the beam life time, and in the optimistic case it should be at least about 4. We assume that the cooling is happening for all 3 degrees of freedom and account that each plane actually makes longitudinal displacement and they need to be summed. Consequently, the cooling ranges should be summed quadratically, and one obtains the total required cooling range $n_\sigma \approx 4\sqrt{3} \approx 7$. That yields the cooling time at the optimal gain of 30 min. Here we additionally accounted a 30% reduction of cooling rate due to reduction of cooling force with momentum deviation which is not accounted in Eq. (56). This correction is much more significant for smaller $n_\sigma$. In particular, for parameters considered in Ref. [35], $n_\sigma$ = 3.7, the accounting of cooling force reduction with amplitude reduces the maximum average cooling rate by almost 3 times. Note that the cooling concept considered in Refs. [35-37] for the microbunched cooling considers one-dimensional cooling model in detail. The obtained cooling time is about 1 hour, *i.e.* the cooler operates well below the optimal gain. That leaves considerable margin for suppression of other diffusion mechanisms not accounted above.

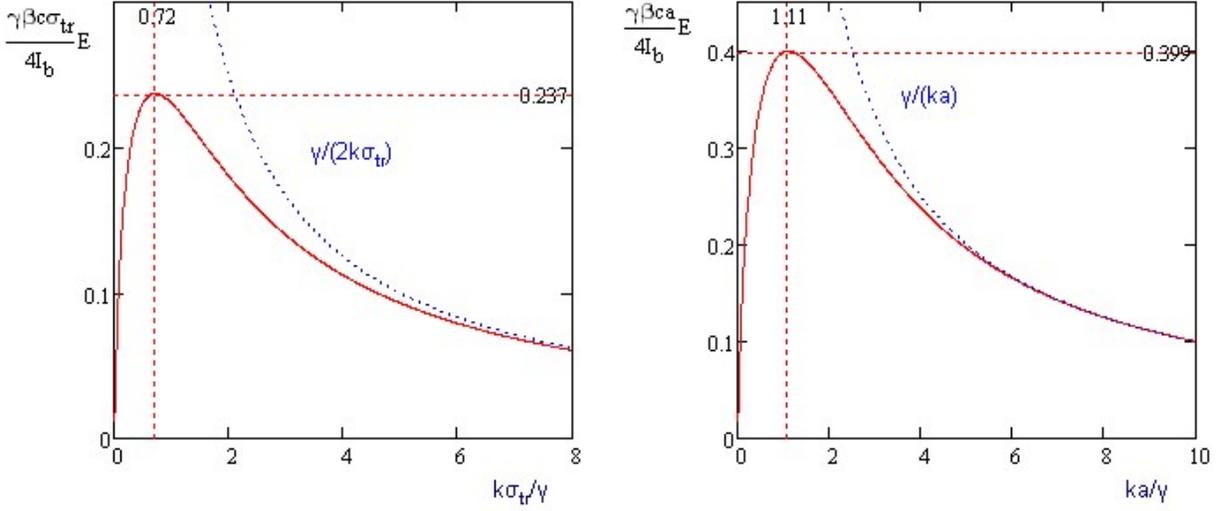

Figure 8: Dependence of dimensionless longitudinal electric field in the beam center, $E/(4\pi e\tilde{n}a)$ on the wave number for Gaussian beam (left) and the round beam with uniform density (right), $b/a_e$=20.

Now we consider a saturation in the CEC amplifier, which represents more severe limitation. The CEC operates on wavelengths $\lambda \leq 2\pi a/\gamma$, where $a$ is the beam radius. At the first step we assume the electron beam density being constant across the beam cross-section. Then, in the beam frame, for the harmonic density perturbation along the bunch $\delta n(r,z,t) = \tilde{n}\exp(-i(\omega t - kz))\Theta(a-r)$, the complex amplitude of longitudinal electric field at the beam axis is expressed by a well-known formula [16]

$$E_z|_{r=0} = \frac{4\pi e\tilde{n}a}{i\gamma}F_E\left(\frac{ka}{\gamma},\frac{b}{a}\right), \quad F_E(x,R) = \frac{1}{x}\left(1 - \frac{I_1(x)K_0(Rx) + I_0(Rx)K_1(x)}{I_0(Rx)(I_0(x)K_1(x) + I_1(x)K_0(x))}\right) \quad (61)$$

where $b$ is the vacuum chamber radius, and $I_0(x)$, $I_1(x)$, $K_0(x)$ and $K_1(x)$ are the Bessel functions. For $b \gg a$ and $ka \geq 1$ the field practically does not depend on $b$. As one can see in the right pane of Figure 8 for a given density perturbation the maximum field is achieved for $ka \approx 1.11$. For $ka \geq 6$ the field is described by an asymptotic corresponding to a single dimensional plasma wave: $E_z|_{r=0} \simeq -4i\pi e\tilde{n}/k$. The numerical solution for the Gaussian distribution was obtained by approximation of the Gaussian distribution by large number of beamlets with uniform density distribution. The results for a Gaussian beam with transverse rms size of $\sigma_{tr}$ are presented in the left pane of Figure 8. For $k\sigma_{tr} \geq 12$ the field is described by an asymptotic corresponding to a single dimensional plasma wave in the beam center. For smaller wave-numbers one needs to account a described above correction related to a 3D structure of electric field. The following fitting formula based on the numerical field computation:

$$E_{zk}\big|_{r=0} \approx -\frac{2I_k}{i\beta c\sigma_{tr}^2} \frac{1}{\sqrt{k^2 + (1.91\gamma/\sigma_{tr})^2}}, \quad \frac{k\sigma_{tr}}{\gamma} \geq 0.72, \tag{62}$$

binds harmonics of electron beam current and its longitudinal field in the beam center. It fits results of numerical computation within few percent for $k\sigma_{tr}/\gamma \geq 0.72$. Note that all proposals for the CEC use the wave-numbers above this boundary. Consequently, accurate accounting for $k\sigma_{tr}/\gamma \leq 0.72$ is not required.

The maximum energy transfer is achieved when the kicker length corresponds to a half of plasma period and a longitudinal proton displacement during this time is much smaller than the wavelength. To account for the finite value of transverse size on the longitudinal kick we use Eq. (62). Then for a Gaussian beam the plasma frequency in the lab frame in the beam center is equal to:

$$\omega_p \approx \frac{1}{\gamma}\sqrt{\frac{4\pi n_0 e^2}{m_e \sqrt{1+(1.91\gamma/(k\sigma_{tr}))^2}}}, \quad \frac{k\sigma_{tr}}{\gamma} \geq 0.72 . \tag{63}$$

Here $n_0 = N_e/((2\pi)^{3/2}\sigma_{tr}^2\sigma_{ze})$ is the unperturbed density in the electron beam center, $\sigma_{ze}$ is its rms length, and $N_e$ is the number of particles in the electron beam. For the beam parameters presented in Table 3 one obtains the length of half plasma period changing in the range of 160 -270 m for $k\sigma_{tr}/\gamma \geq 0.72$, while only 40 m can be allocated for the kicker in RHIC. Consequently, one can neglect the plasma oscillations on the kicker length.

In the CEC the density perturbations excited by different protons overlap resulting in that the total density perturbation is much larger than a perturbation coming from a single proton. Its relative rms fluctuations are:

$$\overline{\delta n_e^2} = \frac{dN_p}{dz}\int_{-\infty}^{\infty} n_1(z)^2 dz , \tag{64}$$

where $n_1(z)$ is the density perturbation excited by one proton, and $dN_p/dz = N_p/\sqrt{2\pi}\sigma_z$ is the peak longitudinal density in the proton beam.

Similar to Ref. [35] we introduce the OSC response to a single proton expressed as a dependence of proton momentum change in the kicker on its longitudinal position, $w(s)$. This function is related to the cooling system gain (introduced in Eq. (17) ) by the following equation;

$$\frac{w(s)}{pc} = \sum_{n=-\infty}^{\infty} G(n\omega_n) e^{2\pi i n \eta_{pk} x}, \quad s = \beta c T_0 \eta_{pk} x \tag{65}$$

where we accounted that $\varepsilon(\omega_n) = 1$ for the system operating at strong band overlap. To find a density perturbation from a known $w(s)$ we use Eq. (62). Then for the relative density (or beam current) perturbation excited by one proton one obtains:

$$\left.\frac{\delta n_1}{n}\right|_k \equiv \left.\frac{\delta I}{I}\right|_k \approx -\frac{i\beta c \sigma_{tr}^2}{2} \sqrt{k^2 + (1.91\gamma/\sigma_{tr})^2} \frac{w_k}{eL_c}, \tag{66}$$

where

$$w_k = \frac{1}{2\pi} \int_{-\infty}^{\infty} w(z) e^{-ikz}. \tag{67}$$

Expressing $\delta n_1(z)$ through its Fourier harmonics, substituting the result into Eq. (64) and performing computations one obtains:

$$\overline{\left(\frac{\delta n_1}{n}\right)^2} = \sqrt{\frac{\pi}{2}} \frac{N_p \sigma_{se}^2 \sigma_{tr}^4}{2 N_e^2 \sigma_s r_e^2 L_k^2 m_e^2 c^4} \int_{-\infty}^{\infty} \left[ \left(\frac{1.91\gamma}{\sigma_{tr}}\right)^2 (w(z))^2 + \left(\frac{dw}{dz}\right)^2 \right] dz \tag{68}$$

Using $w(z)$ presented in the Ref. [35],

$$w(z) = -W_0 \sin\left(2\pi \frac{z}{z_0}\right) \exp\left(-\frac{z^2}{\sigma_0^2}\right),$$

one obtains that two the terms in the integral (68) make close contributions resulting in the total rms density variation of 18%. To effectively operate the dynamic range of stochastic cooling system has to be at least 2.5σ, which results in the maximum density variations of 44%. It does not leave any margin for the noise contribution coming from the shot noise in the electron beam. Simulations in 1D model show that such density perturbations are acceptable. However, it is unclear how it will be changed in a 3D model. There is also a problem with the concept proposed in Ref. [35] where the rms sizes of electron and proton beams are approximately equal. That results in additional reduction in the cooling rates for particles with large betatron amplitudes.

The CEC based on the plasma cascade instability has additional drawbacks. First, this type of cooling requires larger phase advance of plasma oscillations, and therefore its performance is more affected by non-linearity of plasma oscillations appearing at large amplitudes. Second, because there is considerable amplification, the gain length will be shorter than the bunch length resulting in additional reduction of cooling length. This can be mitigated by making the flat top distribution in the electron bunches. However, it is presently unclear how it can be done for the electron beam parameters required for CEC.

Above we considered only longitudinal CEC. The transverse cooling can be considered similar to what is described in the OSC section – an introduction of non-zero dispersion in "pickup" and "kicker" which couples the transverse and longitudinal degrees of freedom. In this case the transfer matrix element $M_{56}$ between pickup and kicker is adjusted to obtain the desired sum of cooling rates while the dispersion is adjusted to obtain the desired ratio of cooling rates. The longitudinal diffusion is not affected by the redistribution of cooling rates. Similar to the IBS [16] the transverse diffusion in the CEC is proportional to the longitudinal diffusion and the dispersion invariant in the "kicker":

$$\frac{d\overline{p_\perp^2}}{dt} = \frac{D^2 + (D'\beta_x + D\alpha_x)^2}{\beta_x} \frac{d\overline{p_\parallel^2}}{dt},$$

where $D$ and $D'$ are the dispersion and its derivative for proton beam in the kicker, and $\beta_x$ and $\alpha_x$ are the corresponding beta- and alpha-functions.

## Conclusions

The beam cooling of proton bunches in the high energy hadron colliders is one of the most challenging problems in the modern accelerator physics. Although considerable progress has been achieved in the recent years there is a number of problems which need to be addressed before a real cooler can be built.

The electron cooling looks as a possible technology for the proton beam energy below ~250-300 GeV. Presently, only ring-based cooling looks realistic for the proton energies above ~100 GeV. With lowering energy, a cooler based on the energy recovery linac looks as a possibility due to reduction of required electron beam current.

The OSC looks as an extremely promising technology for the proton beam energy above ~250-300 GeV. However, it requires the state-of-the-art undulators with field of ~10 T, and an optical amplifier with small signal delay and large gain. In the summer of 2021, the passive OSC was demonstrated in the IOTA ring in Fermilab with 100 MeV electrons [30]. That strongly supports its further development and assures us that the stochastic cooling at the optical frequencies is possible.

The CEC development is still at its initial stage. Although considerable work has been done in recent years its potential and reach need to be understood better before real implementation can be considered.